\documentstyle[12pt,fleqn,aps,psfig]{revtex}

\begin{document}
\draft
\title{Distribution of parallel vortices studied by 
spin-polarized neutron reflectivity and magnetization} 
\author{S.-W. Han} 
\address{Department of Physics
University of Washington, Seattle, Washington 98195}
\author{ P. F. Miceli, J. Farmer}
\address{Department of Physics and Astronomy,
University of Missouri-Columbia, Columbia, Missouri 65211}
\author{G. Felcher, R. Goyette}
\address{Material Science Division, Argonne National Laboratory, Argonne,
Illinois 60439}
\author{G. T. Kiehne, J. B. Ketterson}
\address{Department of Physics and Astronomy and the Materials Research Center,
Northwestern University, Evanston, Illinois 60208}
\date{\today}
\maketitle

\begin{abstract}

We present the studies of non-uniformly distributed vortices 
in Nb/Al multilayers at applied field near parallel to film surface 
by using spin-polarized neutron reflectivity (SPNR) and DC magnetization  
measurements.   
 We have observed peaks above the lower critical field, H$_{c1}$,
 in the {\it M-H} curves from the multilayers. 
 Previous works with a model calculation
of minimizing Gibbs free energy have suggested that
the peaks could be ascribed to vortex line transitions 
for spatial commensuration in a thin film superconductor. 
In order to directly determine the distribution of vortices,
 we performed SPNR measurements on the multilayer  
 and found that the distribution and density 
of vortices are different at ascending and descending fields.
At ascending 2000 Oe which is just below the first peak in the 
{\it M-H} curve, SPNR shows that vortices are mostly localized
near a middle line of the film meanwhile the vortices are distributed 
in broader region at the descending 2000 Oe.  
That is related to the observation of
 more vortices trapped at the descending field. 
As the applied field is sightly tilted ($<$ 3.5$^o$), we observe 
another peak at a smaller field. The peak position is consistent 
with the parallel lower critical field (H$_{c1\parallel}$).
We discuss that the vortices run along the applied
field below H$_{c1\parallel}$ and rotate parallel to the surface  
at H$_{c1\parallel}$.

\end{abstract}
\pacs{PACS numbers: 74.60.Ge, 61.12.Ha, 74.25.Ha}
\vskip 10in
\section{Introduction}

Vortices running parallel to surface in a thin superconducting film
have been widely investigated theoretically and experimentally. 
As the field applied parallel to surface, the Bean-Livingston surface barriers 
significantly contribute to ingress and egress of vortices \cite{bean,degennes}.
They also could do an important role in determining density and 
distribution of vortices, as the film thickness is comparable to the 
London penetration depth \cite{guimpel,brongersma}. 
Unusual prominences in a {\it M-H} curve  
above H$_{c1}$ have been observed from superconducting films by using
several different techniques, including electron tunneling \cite{sutton},
microwave absorption \cite{monceau}, resistivity \cite{yamashita},
superconducting channel device \cite{pruymboom},
SQUID magnetization \cite{guimpel,yusuf}, 
torque magnetization \cite{brongersma}, and vibrating reed
 \cite{hunnekes,ziese}. 
Guimpel {\it et al.} \cite{guimpel} has 
suggested that the peak could be due to the vortex line transitions for 
spatial rearrangements with a model calculation of minimizing Gibbs free 
energy. The idea of minimizing free
energy with vortices has been further developed by Brongersma {\it et al.} 
 with Monte Carlo simulation \cite{brongersma}.
However the above techniques measure only the average magnetization
or the result of vortex motions.  It means that they can not determine 
the location of vortices with the mesurements alone.
Furthermore experimental
measurements from a thin YBa$_2$Cu$_3$O$_{7-x}$ film \cite{hunnekes}
disgree with the model calculation of the free energy with vortices
\cite{ziese}. 
Therefore only a direct measurement on the
location of vortices would clarify whether the peaks comes from the
vortex line transitions. 

Since spin-polarized neutron reflectivity (SPNR) could detects spatial gradient
of magnetic field, it has been used to determine London penetration depth 
 of high-Tc superconductors as well as conventional 
superconductors\cite{felcher} at a small field. 
In the regime of saturated field, SPNR has showed a capability to observe
vortices \cite{han,lauter}.
A theoretical model calculation by Han {\it et al.} \cite{han} has 
demonstrated that it 
could be able to measure the density and also the distribution of vortices.
By using another advantage of SPNR that measures only parallel component of
magnetization due to a scattering selection rule, Han {\it et al.} \cite{han1} 
has demonstrated that vortices placed perpendicular to surface
considerably contribute to the magnetization measurements even at a small
tilted field.

In this paper, we introduce SPNR and magnetization measurements
from the vortices in Nb/Al mulitilayers in Sec. II and present the results 
in Sec. III. A model calculation for minimizing Gibbs free energy
is discussed in Sec. IV and the magnetization measurements
under tilted fields are presented in Sec. V. In the last section,
 we summarize the main conculsions.

\section{Experimental Details}

Nb(72\AA)/Al(20\AA), Nb(100\AA)/Al(20\AA) and Nb(130\AA)/Al(20\AA)
mulilayers with repeating 20 times respectively were deposited on Si 
substrates by direc-current
sputtering under a base pressure of $\sim$10$^{-4}$mTorr and the Ar
partial pressure of 5mTorr. During the deposition, the power was
applied to a Nb target with 275 watts (297 voltages) and a Al target
with 200 watts (372 voltages) at which the deposition rate was 5.9 \AA /sec 
for Nb film and 4.7 \AA /sec for Al film meanwhile the
substrate was placed at the ambient temperature. Tc of
the Nb(72\AA)/Al(20\AA) multilayer, 7.25 $\pm$ 0.25 K, was
determined by magnetization measurement at applied field 50 Oe.

For magnetization measurements, a multilayer mounted on an
extended sample holder was placed in a cryostat with the geometry
where the film surface runs near perpendicular to a pick-up coil of the SQUID
and zero-field cooled. Subsequently the field was applied.
The tilt angle between the surface and  applied field
was controlled by shimming non-magnetic plastic pieces between the sample
holder and sample.
The tilt angle was reproducible within uncertainty, $\pm$0.25$^o$.

SPNR measurements from the 
Nb(72\AA)/Al(20\AA) multilayer were performed at POSY1
reflectometer, Intense Pulsed Neutron Source, Argonne National
Laboratory.
The polarization efficiency $\sim$97\% and the instrumental resolution 
$\Delta$q/q = 0.053 were counted to analyzing the reflectivity data.
The specimen was zero-field cooled and the field was applied
parallel to the film surface at tilt angle $<$ 0.5$^o$.
The polarized neutrons was reflected from the multilayer with 
 incident angle 0.45$^o$.

\section{Data analysis and Results}

Magnetizations from the multilayers were measured
for ascending field at 2 K, as shown in Fig.\ \ref{fig1}. 
We observe the peaks (indicated by small arrows) from the multilayers.
The peak positions are dependent on the configuration of the films.
The peak from the Nb(72\AA)/Al(20\AA) multilayer is more notable than 
the others. The detail magnetization studies will be discussed in Sec. V.
For more understanding of the interfaces of the Nb(72\AA)/Al(20\AA) multilayer, 
specular x-ray reflectivity measurement was carried under atmosphere.
Figure\ \ref{fig2} shows the grazing angle x-ray reflectivity
as a function of momentum transfer, {\it q}.
A best fit (solid line) shows 
Nb(120$\pm$20\AA)/[Al(19$\pm$1.5\AA)/Nb(74.5$\pm$2.5\AA)]$\times$20/Si and
an extra layer at the air/Nb interface with thickness $\sim$60 \AA\
and the x-ray scattering density half of Nb improves a best fit. 
A rms roughness at the air/Nb interface 
is $\sim$15 \AA\  while the roughness of the Si
substrate is $\sim$2.2 \AA. Each Nb layer has the roughness 
$\sim$6 \AA\ and each Al layer has $\sim$2.5 \AA\ rough surface.

SPNR measurements were performed on the Nb[72\AA]/Al[20\AA] multilayer
to directly determine the density and distribution of vortices which run
parallel to the surface. 
Figure\ \ref{fig3} (a) shows neutron specular reflectivities measured 
as a function of {\it q}\ 
for spin up and spin down neutrons at 700 Oe and 2 K.
The oscillation period
is corresponded to the total film thickness $\sim$2020 \AA. A best
fit (solid line) shows that the top Nb layer thickness is 180 $\pm$ 40 \AA\
and multilayer [Al(20 $\pm$ 2\AA)/Nb(72 $\pm$ 5\AA)] $\times$ 20.  
However it could not 
reveal the buried interfaces as well as the x-ray reflectivity does 
because the neutron data were taken in a small-{\it q} region.
The neutron reflectivity is basically consistent with the x-ray 
reflectivity within uncertainty and the configuration of the multilayer 
determined by neutron reflectivity measurement was used for analyzing the 
SPNR data.
Figure\ \ref{fig3} (b) shows reflectivity difference between spin-up and
spin-down devided by their average ($\Delta$R/$\overline R$) that 
more clearly demonstrates magnetization contribution to SPNR. 
The solid line is a best fit without including vortices. It showes that 
the London penetration depth ($\lambda_L$) of the multilayer is 
1800 $\pm$ 200 \AA. 
At a high field, vortices will enter the superconductor and
SPNR would see the vortices. 

Figure\ \ref{fig4} show $\Delta$R/$\overline R$ under 2 K at 
ascending field 1500 Oe (a) which is just above H$_{c1}$ and 
2000 Oe (b) which is just below the first peak, as shown in Fig.\ \ref{fig1}.
Below critical angle ($q$ $\simeq$ 0.013 $\AA^{-1}$), the beam is totally 
reflected by the surface  and the magnetization
contribution to SPNR vanishes. For $q$ $>$ 0.018 $\AA^{-1}$, 
the reflectivity differences among the models are relatively smaller
meanwhile the uncertainty is growing up.  Thus, data in
$q$-range of 0.013 - 0.0175 $\AA^{-1}$ only were fit to 
a theoretical model \cite{han}. 
In the fit, only single parameter, density of vortices, was varying and
the distribution is assumed to be mostly localized
in 1 row (dashed line), 2 rows (solid line) and
 uniformly distributed through the whole specimen (dotted line).

The results of least $\chi$ squares fits with counting the
statistical uncertainty are summarized in table \ref{table1} for
1500 Oe. It shows that
1 row of vortex model is slightly better than the others and the
vortex density is 30 $\pm$ 6 $\mu m^{-2}$ which is corresponding to the
distance of vortex-vortex $\sim$1650 \AA. 
Table \ref{table2} shows the $\chi^2$ of fits for different models
for 2000 Oe.
The models of 1 row and 2 rows are not distinguishable with basing on the
least $\chi^2$. However the magnetic field due to the vortices in a line 
interferes with the surface screening fields
in $\Delta$R/$\overline R$. The interference effect is more
clearly seen near the first peak of $\Delta$R/$\overline R$,
{\it q} $\simeq$ 0.0145 \AA$^{-1}$ in this system.
It suggests that the model of 1 row is better than 2 rows.
From the fit with 1 row (dashed line) the vortex density 
is found to be 45 $\pm$ 6 $\mu m^{-2}$
that is corresponding to $\sim$1100 \AA\ of the average distance between
 two adjacent vortices.

After cycling the field to 5400 Oe, SPNR measurement was also conducted 
from the multilayer at 2000 Oe, as shown in Fig.\ \ref{fig5}.
The lines present a best fit with assuming different distributions of
vortices. A goodness fit of solid line strongly suggests that
the vortices stay in 2 rows, 1/3 and 2/3 of the total film thickness
and the density of vortices 56 $\pm$ 2 $\mu m^{-2}$
The results of fits are summarized in table\ \ref{table3}.
The fits show that 2 rows of vortex  model is about
twice better fit than 1 row and uniform distribution.
From the best fit with 2 rows, the nearest vortex distance
is calculated to be $\sim$1110 \AA\ with assuming
a triangular vortex lattice.
(b) shows the neutron scattering density profile of
 magnetic as well as nuclear potential that
is corresponding to the fit with 2 rows in (a).
The inset shows a vertical expansion.

The best fits of the SPNR data are summarized in table\ \ref{table4}.
The analyses show that vortices stay in a single row at ascending fields, 
1500 and 2000 Oe, however they would like to spread into 2 rows
at the cycled field 2000 Oe. More broadening at the cycled field
could be explained in terms of the trapped vortices
by the surfaces. 
The parallel magnetization (M$\parallel$)
is calculated by using the density and distribution of vortices 
determined by SPNR with a theoretical model\ \cite{han1}.
The M$\parallel$ shows that is unable to distinguish the differences 
among the fields due to large uncertainty. 
However they are certainly negative values even at the cycled field.
We could directly compare those to the SQUID data.
As shown in the table, the magnetizations of M$\parallel$ 
for ascending field are comparable to $\bar {M}$ directly
converted from SQUID data within uncertainty
however there is a big difference at the cycled field.
It suggests that $\bar {M}$ of the cycled field is contributed
by not only M$\parallel$ but also vortices running perpendicular to the
surface. 
That mechanism in which vortices enter the 
superconducting film parallel to the surface for
ascending field and rotated out of plane during reducing the field
has been observed in a Nb superconducting film\ \cite{han1}. 

Although we could obtain reasonable results with the model, 1 row or 2 rows,
 of the vortex distributions, a Gaussian distribution could be an 
alternative choice for the vortices in a superconducting film.
Figure\ \ref{fig6} shows the best fits with a Gaussian distribution of
vortices and table\ \ref{table5} summarizes the results of the fits.
The model of Gaussian distribution absolutely improves the fits, 
particularly 1500 and ascending 2000 Oe. At 1500 and ascending 2000 Oe,
the full width at half maximum (FWHM) is basically the same however
it shows broader at the cycled 2000 Oe. The density and M$\parallel$
are very comparable to the model of 1 row or 2 rows.
Because of a limit of the SPNR sensitivity to this system,
it is not able to distinguish between a Gaussian distribution and 
1 row or 2 rows. However it is clear that the vortices have 
a same distribution at ascending fields and they are differently 
 distributed at the cycled field. 

\section{Model calculation and discussion}

For more theoretical understanding of vortex distribution in a 
superconducting film, we have developed a model calculation for minimizing
Gibbs free energy.
As N number of vortices enter in a superconducting film, 
the total free energy
of the superconductor could be calculated by using a London
approximation \cite{degennes} with simply assuming that 
it is an isotropic superconducting film and impurity in the film is
negligible.
\begin{equation}
G = G_0 + {1 \over 2\mu_o} \int_{V_S} dv \left\{\vec \Phi \cdot (2\vec B_L + 
\vec B_V)\right\}
\label{eq1}
\end{equation}
where G$_0$ is free energy for the system without vortices, $\mu_o$ is 
permeability in vacuum, $V_S$ is volume of superconductor,
$\vec \Phi$ is vorticities, $\vec B_L$ is surface screening field and
$\vec B_V$ is magnetic field due to vortices including their images. 
The vorticities are defined to be,
\begin{equation}
 \vec \Phi  = \Phi_{o} 
\sum_{k = 0}^{k = N} \delta(\vec r - \vec r_k) \hat{x}
\label{eq2}
\end{equation}
where $\Phi_{o}$ is a vortex flux quantum, 2.067 $\times$ 10$^9$ G$\AA^2$,
$\vec r_k$ is location of k$^{th}$ vortex and
vortices are oriented in {\it x}-axis. As assumed that 
the applied field is exponentially decayed from the surface,
the surface screening field is 
\begin{equation}
\vec B_L = \mu_o H \left\{ {cosh(z / \lambda_L) \over cosh(t / 2\lambda_L)}
- 1 \right\} \hat{x}
\label{eq3}
\end{equation}
where {\it t} is film thickness and applied field is along {\it x}-axis. 
If the vortices are localized in lines,
$\vec r_k$ = z$_k$ $\hat{z}$ + {\it kl} $\hat{y}$ in Eq.\ (\ref{eq2}) 
where {\it l} is average distance of adjacent vortices in $\hat{y}$-direction
and the the spatial magnetic field due to the vortices will be
\begin{equation}
 \vec B_V = {\Phi_o \over 2\pi \lambda_L^2} \sum_{k=-N/2}^{k=N/2}
\sum_{n=-\infty}^{n=\infty} (-1)^n K_0 \left\{{\sqrt{(z - nt - (-1)^nz_k)^2
+(y - kl)^2)} \over \lambda_L} \right\} \hat{x}
\label{eq4}
\end{equation}
where $K_0$ is a modified Bessel function of the first order. 
For simplifying the free energy calculation, previous studies 
\cite{brongersma,gilson} have used 
an approximation which is valid only for a limit of $\lambda_L$ $\gg$ a 
film thickness. Since the approximation is not 
applicable for
our system ($\lambda_L$ $<$ {\it t}), we have to count term by term in 
the summation of Eq.\ (\ref{eq4}).

For the free energy calculation, we need two characteristic lengths,
London penetration depth and coherence length. 
$\lambda_L$ $\simeq$ 1800 \AA\ for the Nb(72\AA)/Al(20\AA)
multilayer was determiend by SPNR. 
The coherence lenght could be obtained by measuring H$_{c1}$.
As the field applied parallel to the surface, the low critical field  
can be estimated by using the London theory \cite{degennes,abrikosov}.
With assuming that a vortex first enters a thin superconducting film 
as the free energy is zero at {\it z} where the vortex is placed, the
lower critical field is
\[H_{c1} = {\Phi_o \over 2\pi\lambda_L^2}
{1 \over 1 - cosh(z / \lambda_L) / cosh(t / 2 \lambda_L)} \]
\begin{equation}
\times \left\{K_0\left({\xi \over
\lambda_L}\right)
+ \sum_{n=1}^{\infty}(-1)^{n} K_0\left(|z - nt - (-1)^nz| \over 
\lambda_L\right) \right\}
\label{eq5}
\end{equation}
We assum that vortices first enter the superconductor, 
as the free energy is zero at the surface, e.q., 
z = t / 2 - $\xi$ in Eq.\ (\ref{eq5}) \cite{han2}.
Based on the magnetization measurement, H$_{c1}$ = 1200 $\pm$ 200 Oe 
and $\xi$ is calculated to be $\sim$113 \AA. 
Although the Nb(72\AA)/Al(20)\AA\ multilayer is an anisotropic superconductor,
an anisotropy of a Nb(100\AA)/Cu(100\AA)
multilayer, $\xi_y$/$\xi_z$, was 1.23 \cite{brongersma}, where $\xi_y$ and
$\xi_z$ are the coherence lengths of in-plane and out of plane respectively. 
 As comparing configuration of
our specimen to Brongermas's one, the asumption of an isotropic superconductor
for the calculation would not be seriously wrong.

Fig.\ \ref{fig7} shows the minimum free energy calculation. 
For the calculation, it is assumed that vortices are localized
in a central line ({\it t} / 2) (dashed line) and two lines ({\it t} / 3 
and 2{\it t} / 3) (solid line), as shown in the inset.
At a given field, the minimum free energy was determined by varying 
the density of vortices only. 
At small fields, the free energy of the system for 1 row is
smaller than for 2 rows whereas above 2200 Oe it is reverse. 
It means that the vortices more likely stay in 2 rows than 1 row
for  H $>$ 2200 Oe. 
This calculation agrees well with the first peak in the magnetization 
measurement, as shown in Fig.\ \ref{fig1} top and strongly suggests that 
the peak is due to a vortex line transition from 1 row to 2 rows.
The inset at upper right corner shows magnetization which corresponds to 
the minimum free energy.
There is the second transition (2 $\to$ 3) at $\sim$4000 Oe
that is also consistent with the second peak in the {\it M-H} curve. 

The calculation shows that vortex density of 40 $\mu m^{-2}$ at 1500 Oe 
and 53 $\mu m^{-2}$
at 2000 Oe will satisfy the condition of minimizing the free energy with
1 row of vortices in this system. 
From SPNR measurements, it is found to be $\sim$30 $\mu m^{-2}$
at 1500 Oe, $\sim$45 $\mu m^{-2}$ at 2000 Oe and $\sim$56 $\mu m^{-2}$
 at cycled 2000 Oe. 
The smaller vortex densities at ascending fields can be understood
in terms of the surface barriers because the minimizing free energy 
calculation does not count the barriers.
Since the vortex densities at ascending fields are smaller than the maximum
vortex density for 1 row, the vortices could stay in a central line 
however at the cycled field, the vortex density found by SPNR is slightly
higher than the the maximum vortex density for 1 row. Thus they can not 
stay in a single row. It means that the vortex line transition fields can be
shifted to a lower field for descending field, as the contribution 
of the surface barrier is important. It has been experimentally observed by
J. Sutton {\it et al.} \cite{sutton}. 

\section{Magnetization at small tilted field}

As the field is tilted with a small angle, the delay of vortex entrance due to
the surface barrier vanishes because the perpendicular component of 
applied field helps the vortex ingress at a small field. 
Figure\ \ref{fig8} shows the magnetization at different tilt angles.
There is a peak at H $\simeq$ 2250 Oe from Nb(72\AA)/Al(20\AA) film
 without depending on the tilt angle ($\theta$ $<$ 3.5$^o$). However
we find another peak at $\sim$950 Oe of the tilted field. It is the first
observation that the first peak is missing at no-tilted field.
One could easily overlook that the first peak does not come from
the vortex line transition. 
Since it is a thin film, vortices can not enter
the superconductor until the applied field is stronger than the surface
barrier field. At a tilted field, however,
vortices can overcome the surface barrier under support by the perpendicular 
component of the field even with a small field. Therefore the vortex line
transitions could be missing at no-tilted field.
This scenario could work only for that the surface barrier field is larger
than the vortex line transition fields. That is not this case because
the measured surface barrier
field for the specimen is about 1200 Oe and SPNR measurement 
shows 1 row of vortices at ascending fields of 1500 and 2000 Oe. 
Also the calculation for minimizing the free energy suggests that 
the first vortex line transition may occur at $\sim$2200 Oe. 
Thus, we suspect that the first peak at the tilted field comes 
a vortex line transition. 

Using Eq.\ (\ref{eq5}),  H$_{c1}$ of this system under the equilibrium
condition where vortices first enter
the superconducting film as the free energy is zero
at the middle of the film, e.q., {\it z} = 0 in Eq.\ (\ref{eq5}), is 
calculated to be
$\sim$850 Oe. That is very comparable to the first peak at the tilted field. 
It might suggest that the peak is due to the lower critical field which is
paralle to the surface (H$_{c1\parallel}$). 
At a tilted field ($\theta >$ 1$^o$ for this system), the first vortex
entrance is determined by H$_{c1\perp}$ instead of H$_{c1\parallel}$
because for a thin film superconductor, H$_{c1\perp}$ is much smaller than 
H$_{c1\parallel}$ \cite{bulaevskii}.
For H$sin\theta$ (H$_{\perp}$) $>$ H$_{c1\perp}$, vortices could enter
the superconductor and might stay along the applied field due
to the dragging force from the applied field. However the free energy still
does not allow the vortices running parallel to the surfaces.
When the applied field (H$cos\theta$) is increased with passing beyond  
H$_{c1\parallel}$, vortices sightly rotate to parallel to the surfaces, 
 connect the pieces, and make long threads. 
In this scenario, however, 
the magnetization might be just a little changed because the field is 
applied with a small tilt angle. 
However we find that a large demagnetization effect enhances the change by 
more than two orders of magnitude because of the thin film geometry. 
The demagnetization of the multilayers is discussed in detail
below.

Figure\ \ref{fig8} (b) and (c) show the magnetization from the
 multilayers at 4.5 K.  The magnetization from 
Nb(100\AA)/Al(20\AA) also shows that the first 
peak is missing at no-tilted field whereas it is observed at 600 Oe
of a tilted field. 
That agrees well with H$_{c1\parallel}$ = 600 Oe calculated by
using Eq.\ (\ref{eq5}) (z = 0), and $\lambda_L$ = 1800 \AA\ and $\xi$ = 113 \AA.
The second and third observed peaks, $\sim$1400 and $\sim$2450 Oe,
can also be compared to the vortex line transitions, 1$\to$2 (1450 Oe) and 
2$\to3$ (2650 Oe) respectively estimated by the model calculation.
That is another evidence that the first peak at a tilted field comes
from the vortex rotation instead of the vortex line transition.
At 4.5 K, two peaks are observed at 845 and 2125 Oe from 
Nb(72\AA)/Al(20\AA). 
The first peak near 845 Oe appears without depending on the tilt angle. 
It implies that the superconductor is soft and the surface barrier
is negligible at the temperature.  Since the peak positions are
not very sensitive to the temperature that is consistent with previous
measurements \cite{brongersma,yusuf}, we might also conclude that
the first peak comes from the vortex rotation at H$_{c1\parallel}$.
The peaks from the Nb(72\AA)/Al(20\AA) and Nb(100\AA)/Al(20\AA) multilayers
were carefully measured  at different temperatures 
and tilt angles and summarized in table\ \ref{table5}.
Table\ \ref{table5} shows that the vortex line transitions (2nd and 
3rd peaks) occur at smaller fields for a thicker film superconductor
because vortices have relatively more space to the surfaces
in a thicker film and rearrange for spatial commensuration
at a smaller field \cite{guimpel,sutton,hunnekes}.

Demagnetization factors of the multilayers are quantitatively
determined in the Meissner regime.
In the regime, slopes of the magnetization were measured
as a function of tilt angle, as shown in Fig.\ \ref{fig9}.
The solid lines are a best fit to data with a model for the
demagnetization of a thin superconducting film \cite{han1}.
The demagnetization factors of Nb(72\AA)/Al(20\AA) are found
to be 0.9986 $\pm$ 0.0011 at 2 K and 0.9935 $\pm$ 0.0007 at 4.5 K while
it is 0.992 $\pm$ 0.0034 for Nb(100\AA)/Al(20\AA) at 4.5 K.
Those can be compared to theoretical calculations, 0.993, 0.9942 
and 0.9934 respectively for our sample geometries\cite{craik}. 
For H $<$ H$_{c1\parallel}$, the perpendicular magnetic field due to 
vortices 
is $n\Phi_{eff}sin\theta/(1 - N)$ where {\it n} is number of vortices per 
unit area, $\Phi_{0eff}$ is the effective flux due to the images and 
{\it N} is the demagnetization factor. 
As the vortices rotate parallel to the surface at H$_{c1\parallel}$,
the perpendicular magnetic field will vanish. As assumed that a single
vortex per $\mu m^2$ rotates parallel to the surfaces at the tilt angle 
2.5$^o$, the difference of a few Gauss for our sample geometry 
contributes to the SQUID magnetization measurement.

\section{Conclusions}

Vortex pinning is an important subject for practical application
of current transportation. With only thin film geometry where vortices
place parallel to the surfaces, the vortex-surface interaction could be
studied. We performed SPNR measurements on a Nb/Al multilayer to 
directly determine the density and distribution of vortices.
SPNR shows that the vortices in a Nb/Al multilayer are localized
near the central line below the first peak in the {\it M-H} curve
at the ascending field meanwhile vortices have more broadening of
the distribution at the descending field. The SPNR measurement is
consistent with the magnetization measurement and also
a model calculation of minimizing Gibbs free energy.
As the field is applied with an angle, another peak which
is missing at no-tilted field is first time observed. Comparing the peak 
to a model calculation of H$_{c1\parallel}$, we conclude that the first
peak at a tilted field comes from the vortex rotation instead of vortex 
line transition at H$_{c1\parallel}$.

Support is gratefully acknowledged (PFM, SWH) from the Midwest
Superconductivity Consortium (MISCON) under DOE grant DE-FG02-90ER45427,
the NSF DMR 96-23827, and the use of the Intense Pulsed Neutron Source
at Argonne National Laboratory. This facility is funded by the U. S. Department
of Energy, BES-Materials Sciences, under Contract W-31-109-Eng-38.

\begin{figure}
\vskip 0.3in
\caption{shows magnetization measured from  Nb/Al multilayers
at 2 K. The data of Nb(100 \AA)/Al(20 \AA) and Nb(130 \AA)/Al(20 \AA) 
were shifted down along the y axis for clarity.
The small arrows indicate the peaks and the big arrows points the
direction of the measurements.}
\label{fig1}
\end{figure}

\begin{figure}
\vskip 0.3in
\caption{X-ray reflectivity was measured from the Nb(72\AA)/Al(20\AA) 
multilayer at atmosphere.
Dotted line is data and solid line is a best fit.} 
\label{fig2}
\end{figure}

\begin{figure}
\vskip 0.3in
\caption{(a) shows grazing angle reflectivities for spin up and spin
down neutrons from a Nb(72\AA)/Al(20\AA) multilayer measured as a function of
{\it q} at 700 Oe and 2 K. The solid line is a best fit. 
(b) shows $\Delta$R/$\overline R$ (described in the text) 
obtained from the data in (a).
The solid line is a best fit and shows that the London penetration depth of 
the multilayer is $\sim$1800 \AA.}
\label{fig3}
\end{figure}

\begin{figure}
\vskip 0.3in
\caption{shows $\Delta$R/$\overline R$ measured
 from a Nb(72\AA)/Al(20\AA) multilayer at ascending 1500 Oe (a) and 2000 
Oe (b) under 2 K.
The lines are a best fit with assuming different distributions of vortices.}
\label{fig4}
\end{figure}

\begin{figure}
\vskip 0.3in
\caption{(a) shows $\Delta$R/$\overline R$ measured from a Nb(72\AA)/Al(20\AA)
multilayer at cycled field 2000 Oe under 2 K.
 The lines are a best fit with assuming different
distributions of vortices.
(b) shows the neutron scattering density profile
that is corresponding to 2 rows 
 of vortices in (a).  The inset is a vertical expansion.}
\label{fig5}
\end{figure}

\begin{figure}
\vskip 0.3in
\caption{shows $\Delta$R/$\overline R$ at 1500, 2000 and (cycled) 2000
Oe and the solid lines are a best fit with a Gaussian distribution
of vortices.} 
\label{fig6}
\end{figure}

\begin{figure}
\vskip 0.3in
\caption{shows minimum Gibbs free energy as a function of applied field
with assuming 1 row (dashed line) and 2 rows (solid line) of vortices.
The assumed locations of vortices are shown in the inset at lower left.
The inset at upper right shows the magnetization corresponding to
the minimum free energy.}
\label{fig7}  
\end{figure}

\begin{figure}
\vskip 0.3in
\caption{Magnetic moments were measured from Nb/Al multilayers 
at different tilt angles and different temperatures. The arrows in (a) 
indicate the direction of measurement. Data in (b) and (c) were taken along
 the same direction of measurement in (a). The lines are a guide to the eye.} 
\label{fig8}
\end{figure}

\begin{figure}
\vskip 0.3in
\caption{shows the slopes of magnetic moment below H$_{c1}$ as
a function of tilt angle. The solid lines are a best fit with a
model [15]. The data (open circle and solid triangle) are shifted 
down along the y axis for clarity.}
\label{fig9}
\end{figure}

\begin{table}
\caption{Least-$\chi^2$ fit at 1500 Oe}
\begin{tabular}{cccc}
& 1 row & 2 rows &  Uniform \\
\tableline
$\chi^2$: & 2.788 (30 $\mu$m$^{-2}$)
 & 2.974 (30 $\mu$m$^{-2}$) & 3.499 (35 $\mu$m$^{-2}$) \\
\end{tabular}
\label{table1}
\end{table}

\begin{table}
\caption{Least-$\chi^2$ fit at 2000 Oe}
\begin{tabular}{cccc}
& 1 row & 2 rows & Uniform  \\
\tableline
$\chi^2$: & 2.945 (45 $\mu$m$^{-2}$) & 2.976 (45 $\mu$m$^{-2}$) 
 & 3.084 (60 $\mu$m$^{-2}$) \\
\end{tabular}
\label{table2}
\end{table}

\begin{table}
\caption{Least-$\chi^2$ fit at cycled 2000 Oe}
\begin{tabular}{cccc}
 & 1 row & 2 rows & Uniform \\
\tableline
$\chi^2$: & 1.855 (50 $\mu$m$^{-2}$)
 & 0.969 (56 $\mu$m$^{-2}$) & 1.498 (72 $\mu$m$^{-2}$) \\
\end{tabular}
\label{table3}
\end{table}

\begin{table}
\caption{Results of SPNR measurements}
\begin{tabular}{ccccc}
H (Oe) &Distribution & Density per $\mu m^2$ 
& M$_\parallel$ (G) 
& $\bar M$ (G)\tablenote{Magnetization directly converted from SQUID data 
in Fig.\ \ref{fig8} (a) open circle with no adjustable parameter.} \\
\tableline
1500 & 1 row & 30 $\pm$ 6 & -51 $\pm$ 17 & -66.6 $\pm$ 7 \\
2000 & 1 row & 45 $\pm$ 6 & -57 $\pm$ 17 & -62.8 $\pm$ 6 \\
2000(cycled) & 2 rows & 56 $\pm$ 2  & -42.5 $\pm$ 5 & 83 $\pm$ 8 \\
\end{tabular}
\label{table4}
\end{table}

\begin{table}
\caption{Least-$\chi^2$ fit with a Gaussian distribution of vortices} 
\begin{tabular}{ccccc}
H (Oe) & $\chi^2$ & FWHM (\AA) & vortex density ($\mu m^{-2}$) & 
M$_\parallel$ (G) \\
\tableline
1500 & 2.056 & 510 $\pm$ 51 & 33 $\pm$ 3 & -47 $\pm$ 9 \\ 
2000 & 2.077 & 530 $\pm$ 29 & 47 $\pm$ 3 & -58 $\pm$ 9 \\  
2000 (cycled) & 0.774 & 700 $\pm$ 23 & 55 $\pm$ 2 & -42 $\pm$ 6 \\  
\end{tabular}
\label{table3-1}
\end{table}

\begin{table}
\caption{Peak positions(Oe)}
\begin{tabular}{ccccc}
Configuration of multilayer &1st &2nd &3rd \\
\tableline
Nb/[Al(20\AA)/Nb(72\AA)]$\times$20 & 950(2K, 4.5K) & 2250(2K, 4.5K)& 3900(2K)\\
Nb/[Al(20\AA)/Nb(100\AA)]$\times$20 & 650(4.5K) & 1400(2K, 4.5K) & 2450(2K, 4.5K)\\
\end{tabular}
\label{table5}
\end{table}


\begin{references}
\bibitem{bean}C. P. Bean and J. D. Livingston, Phys. Rev. Lett. {\bf 12}, 
14 (1964).
\bibitem{degennes}P. G. de Gennes, {\it Superconductivity of Metals and
Alloys} (Addison-Wesley, New York, 1989); T. P. Orlando, K. A.
Delin, {\it Foundations of Applied Superconductivity} (Addison-Wesley,
New York, 1991).
\bibitem{guimpel}J. Guimpel, L. Civale, F. de la Cruz, J. M. Murduck, I. K.
Schuller, Phys. Rev. B {\bf 38}, 2342 (1988).
\bibitem{brongersma}S. H. Brongersma, E. Verweij, N. J. Koeman, D. G. de Groot,
R. Griessen, B. I. Ivlev, Phys. Rev. Lett. {\bf 71}, 3219 (1993).
\bibitem{sutton}J. Sutton, Proc. Phys. Soc. {\bf 87}, 791 (1966).
\bibitem{monceau} P. Monceau, D. Saint-James, G. Waysand, Phys. Rev. B
{\bf 12}, 3673 (1975).
\bibitem{yamashita} T. Yamashita, L. Rinderer, J. Low Temp. Phys. {\bf 
24}, 695 (1976), N. Ya. Fogel, V. G. Cherkasova, Physica B {\bf 107},
291 (1981), P. Lobotka, I. V\'{a}vra, R. Sender\'{a}k, D. Machajd\'{i}k,
M. Jergel, \v{S}. Ga\v{z}i, E. Rosseel, M. Baert, Y. Bruynseraede, M.
Forsthuber, G. Hilscher, Physica C {\bf 299}, 231 (1994). 
\bibitem{pruymboom} A. Pruymboom, P. H. Kes, E. van der Drift, S. Radelaar,
Phys. Rev. Lett. {\bf 60}, 1430 (1988).  
\bibitem{yusuf} S. M. Yusuf, E. E. Fullerton, R. M. Osgood II, G. P. Felcher,
J. appl. Phys. {\bf 83}, 6801 (1998).
\bibitem{hunnekes} G. H\"unnekes, H. G. Bohn, W. Schilling, H. Schulz,
Phys. Rev. Lett. {\bf 72}, 2271 (1994).
\bibitem{ziese} M. Ziese P. Esquinazi, P. Wagner, H. Adrian, S. H. Brongersma,
R. Griessen, Phys. Rev. B {\bf 53}, 8658 (1996).
\bibitem{felcher}G. P. Felcher, R. T. Kampwirth, K. E. Gray, R. Felici,
Phys. Rev. Lett. {\bf 52}, 1539 (1984); H. Zhang, J. W. Lynn, C. F. Majkrzak,
S. K. Satija, J. H. Kang, X. D. Wu, Phys. Rev. B {\bf 52}, 10395 (1995);
A. Mansour, R. O. Hilleke, G. P. Felcher,
R. B. Lainbowitz, P.  Chaudhari, S. S. P. Parkin Physica B {\bf 156 \&
157}, 867 (1989); S. V. Gaponov, E. B. Dokukin, D. A. Korneev, 
E. B. Klyuenkov, W. L\"{o}bner, V. V. Pasyuk, A. V. Petrenko, Kh. Rzhany,
and L. P. Chernenko, JETP Lett. {\bf 49}, 316 (1989); V. Lauter-Pasyuk,
H. J. Aksenov, E. L. Kornilov, A. V. Petrenko, P. Leiderer, Physica B
{\bf 248} 166 (1998).
\bibitem{han} S.-W. Han, J. F. Ankner, H. Kaiser, E. Paraoanu, L. H. Greene,
P. F. Miceli, Phys. Rev. B {\bf 59}, 14 692 (1999).
\bibitem{lauter} V. Lauter-Pasyuk, H. J. Lauter, M. Lorenz, V. L. Aksenov,
and P. Leiderer, Physica B {\bf 267-268}, 149 (1999).
\bibitem{han1}S.-W. Han, J. Farmer, P.F. Miceli, I. R. Roshchin, L. H. Greene,
Phys. Rev. B {\bf 62}, 9784 (2000).
\bibitem{abrikosov}A. A. Abrikosov, {\it Fundamentals of the theory of metals}
(North-Holland, 1988).
\bibitem{han2}S.-W. Han, Ph. D. Thesis, Department of Physics (University
of Missouri, 1999), S.-W. Han and P.F. Miceli (unpublished).
\bibitem{gilson} Gilson Carneiro, Phys. Rev. B {\bf 57}, 6077 (1998),
Y. Mawatari, K. Yamafuji, Physica C {\bf 228}, 336 (1994).
\bibitem{bulaevskii} L. N. Bulaevskii, M. Ledvij, and V. G. Kogan, Phys.
Rev. B {\bf 46}, 366 (1992).
\bibitem{craik}D. Craik, {\it Magnetism Principles and Applications} (Wiley,
New York, 1995) p. 298.
\end{references}
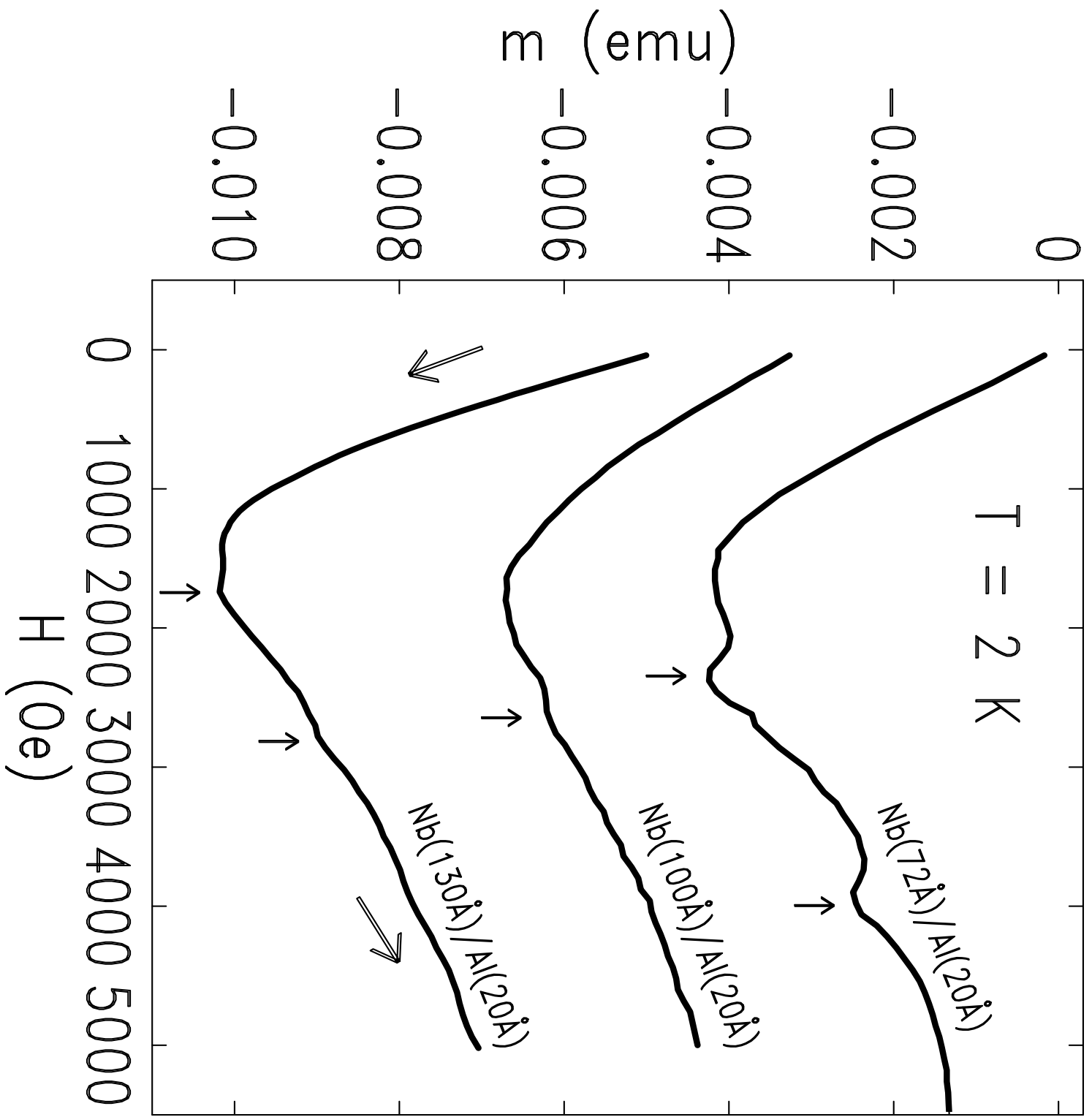
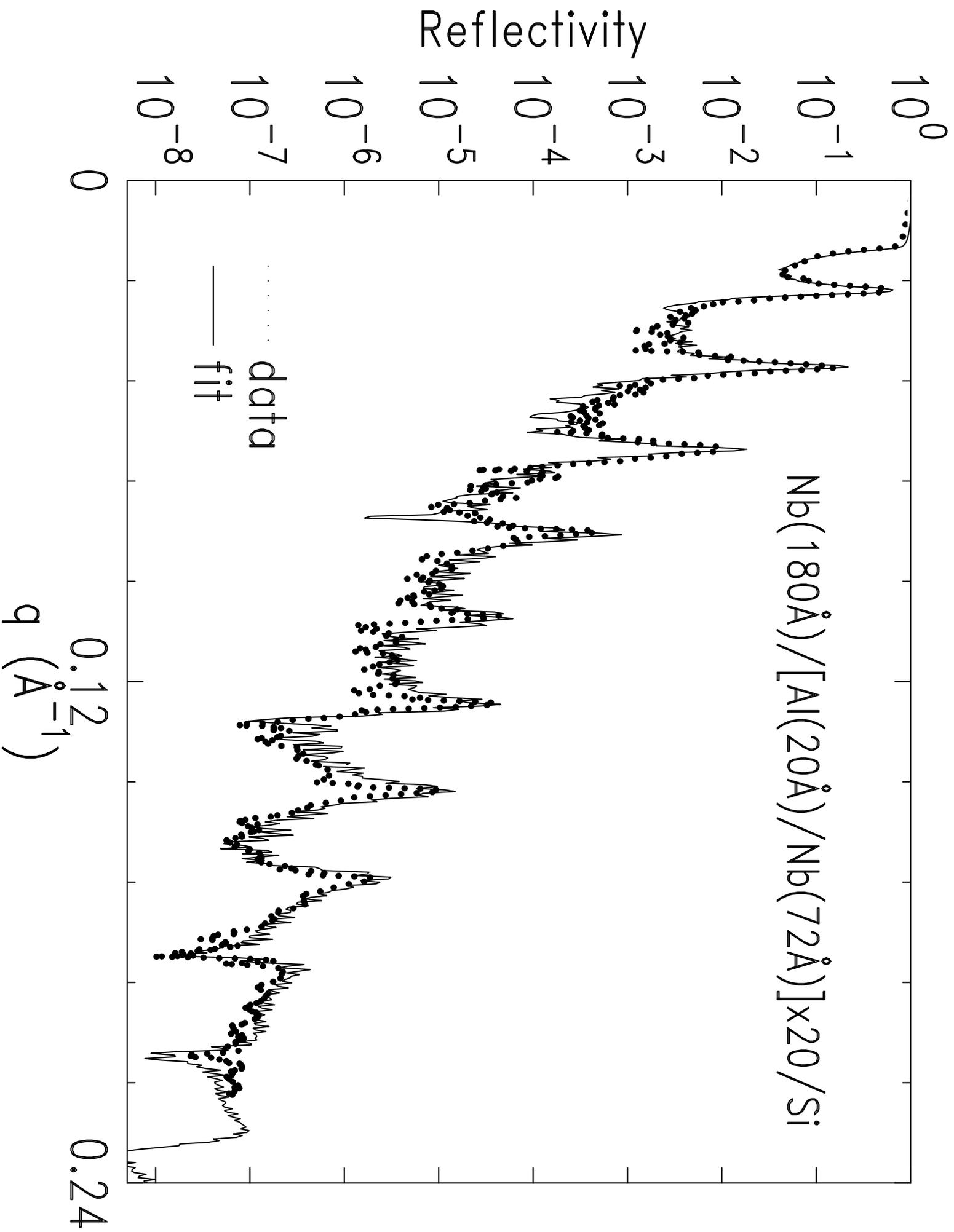
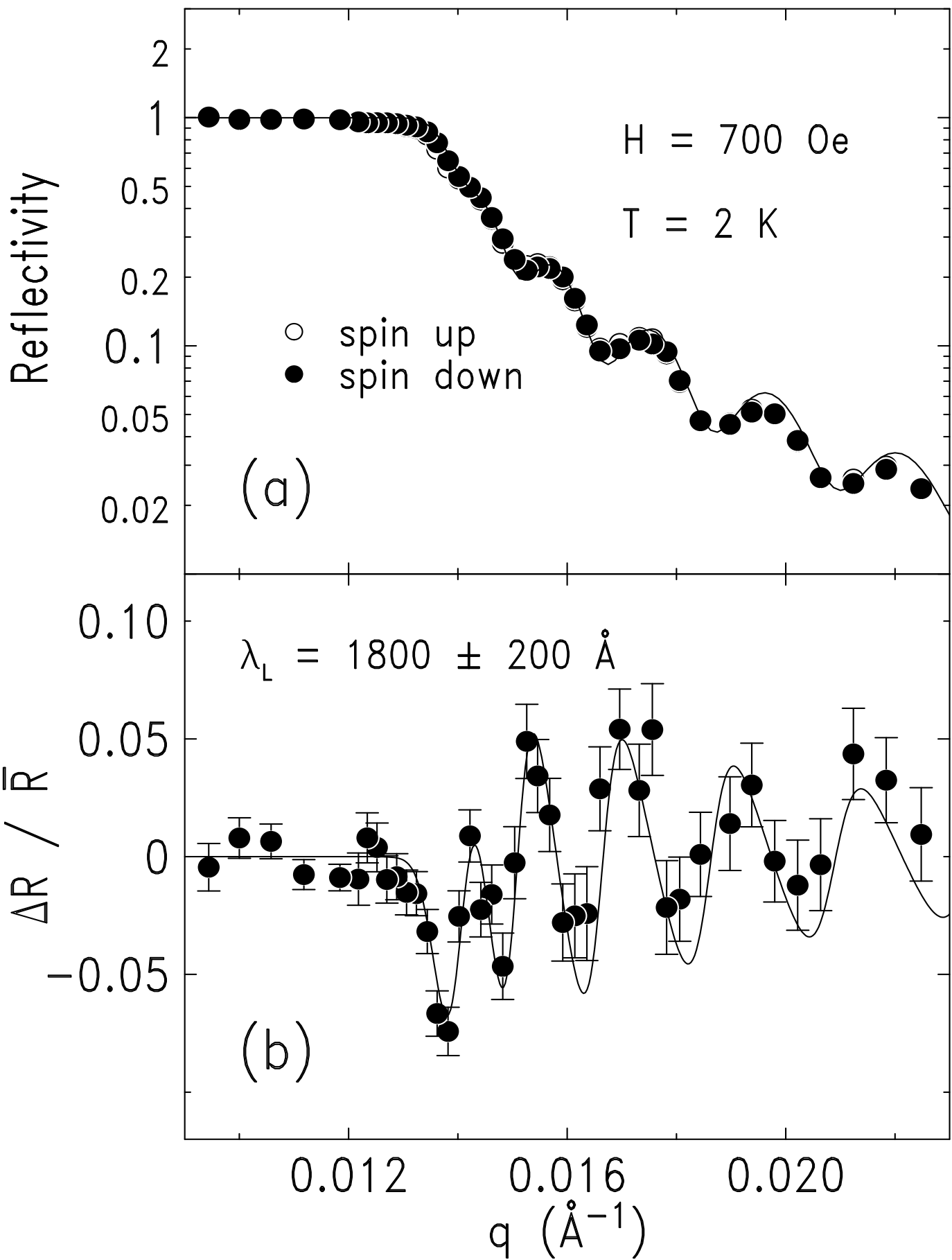
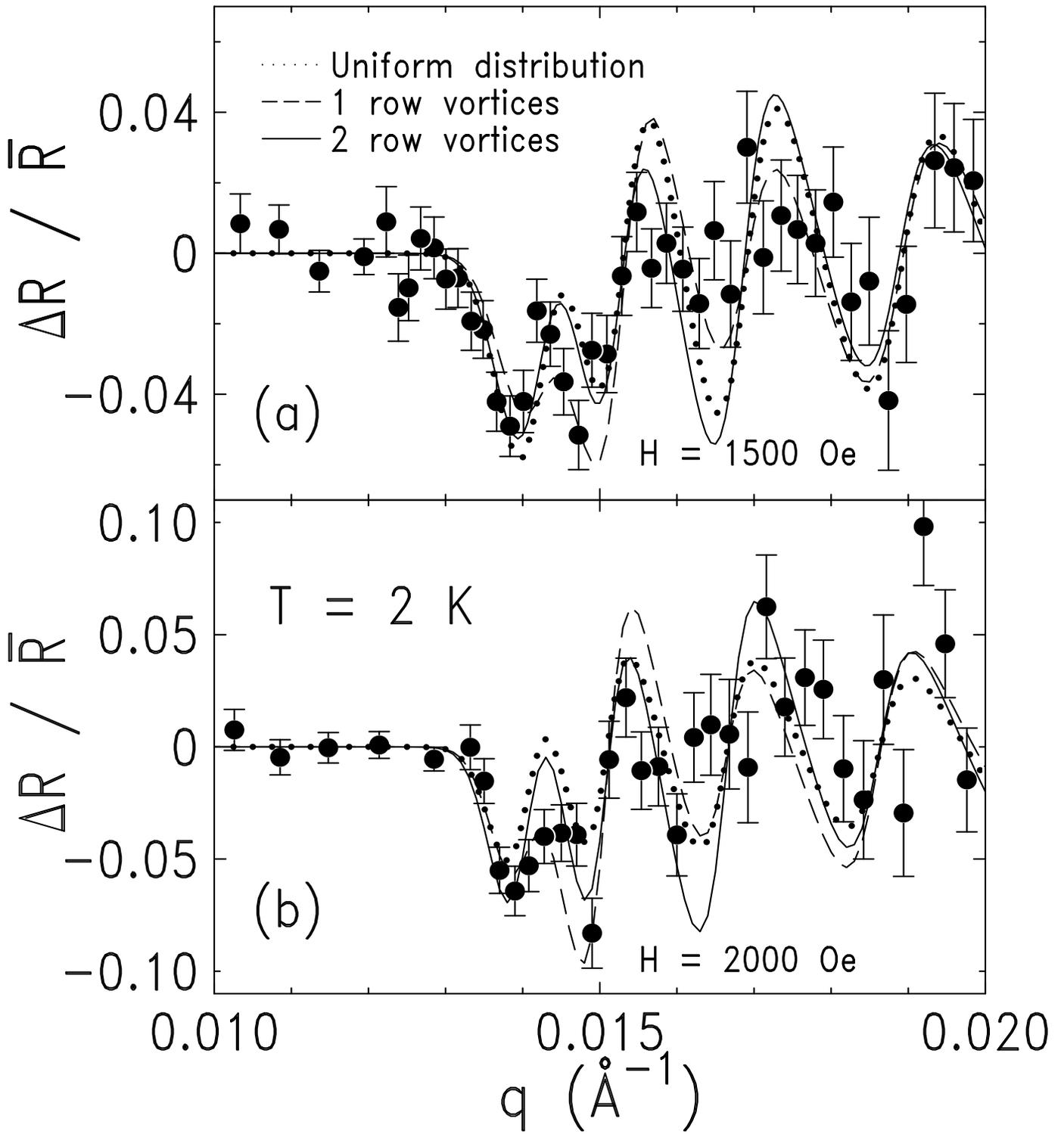
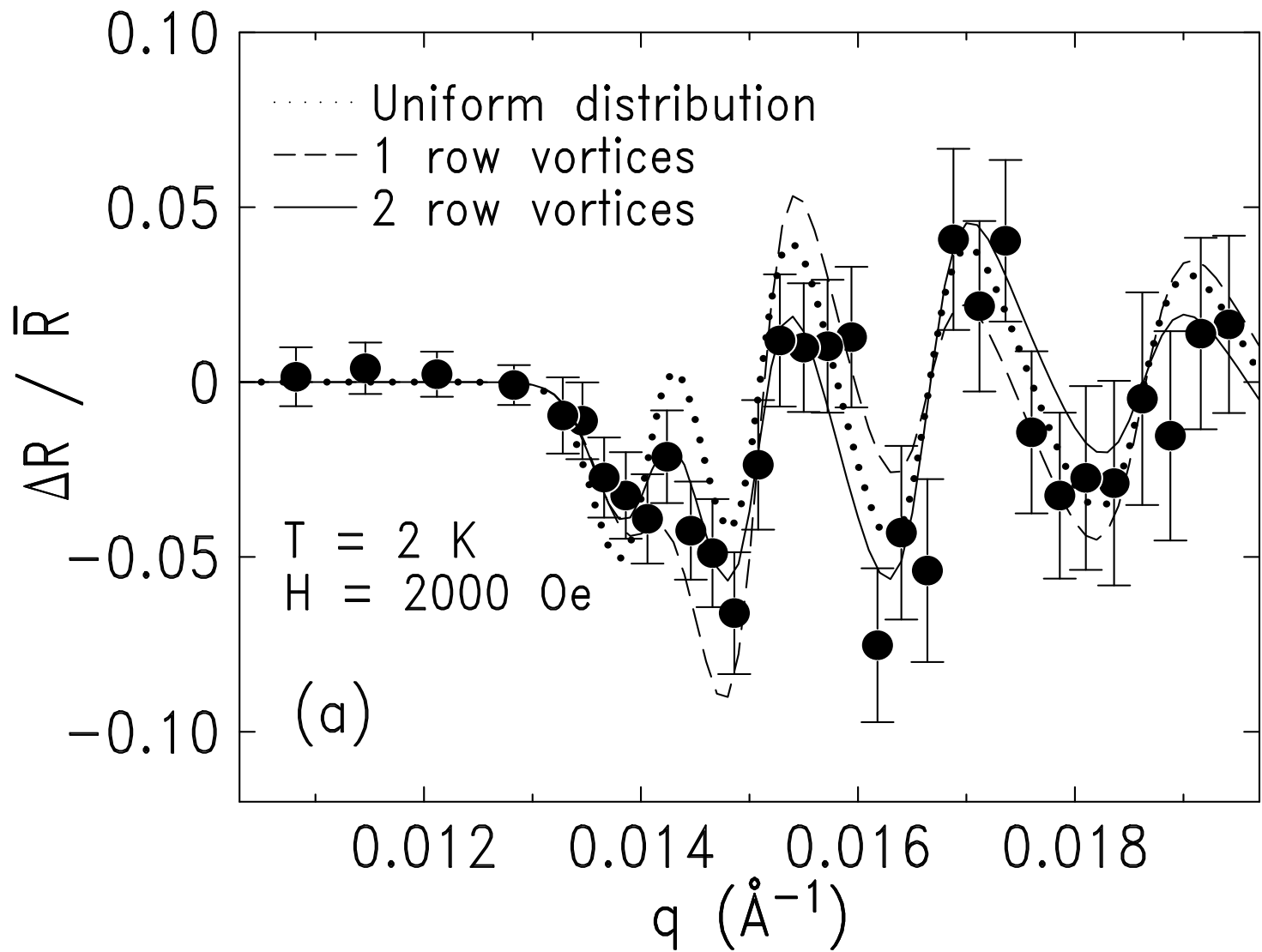
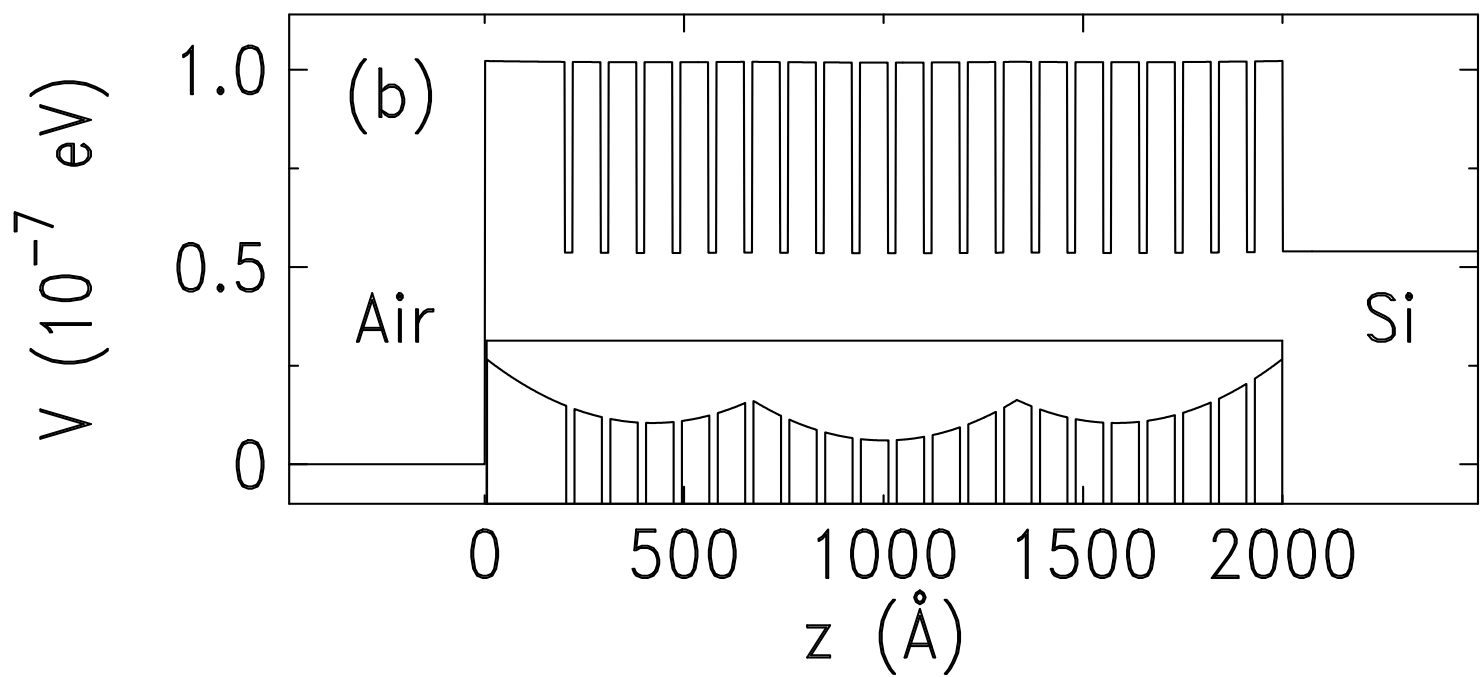
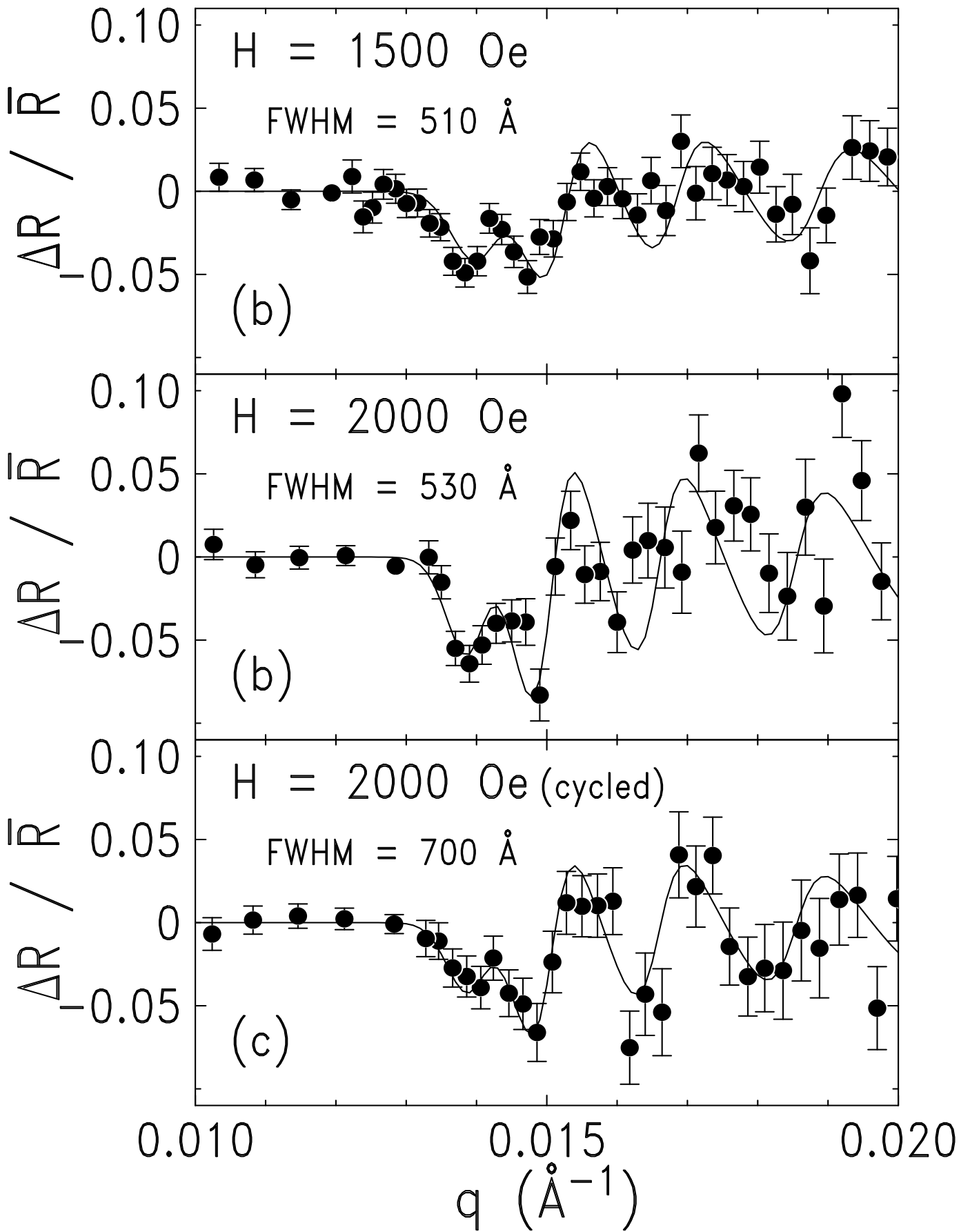
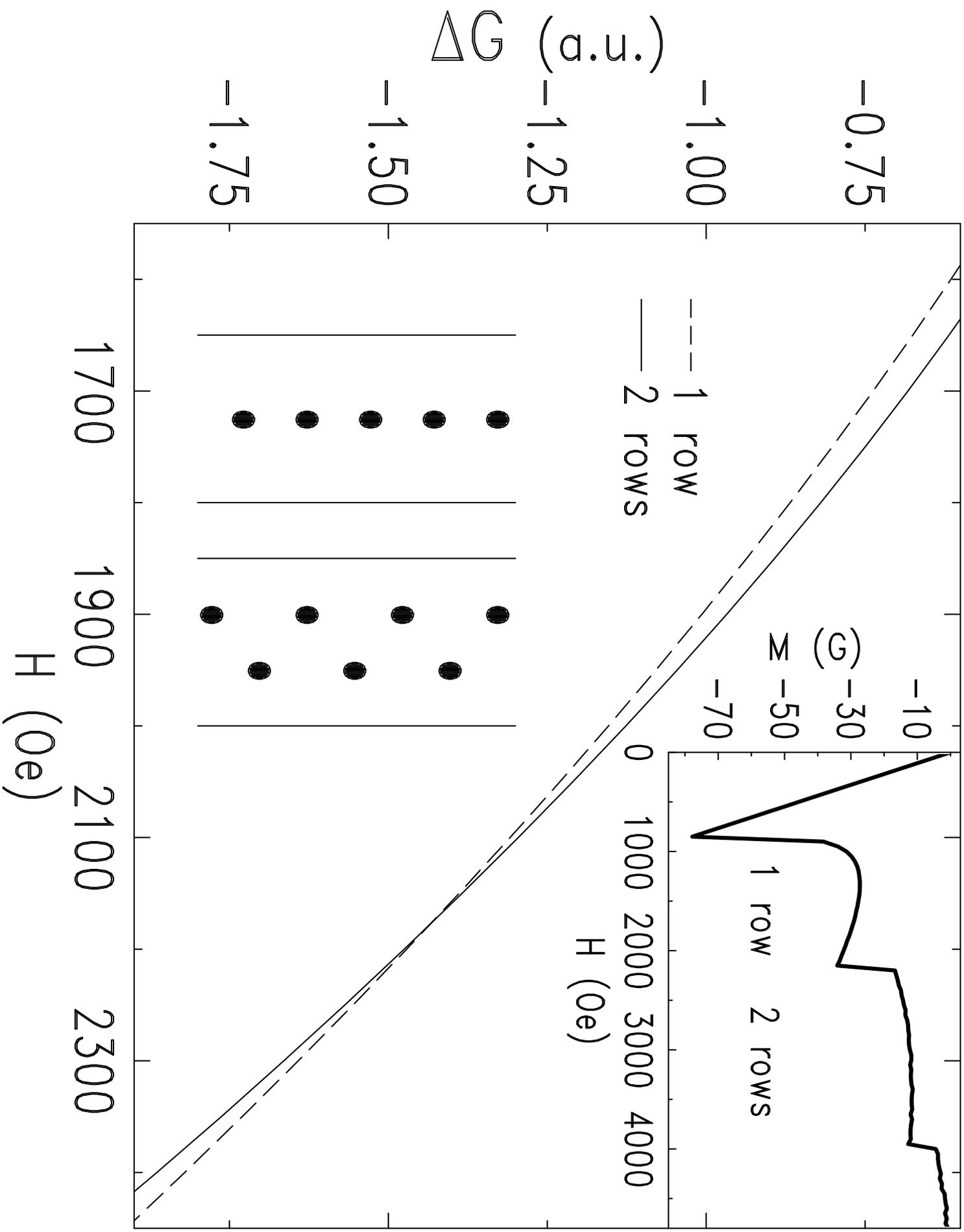
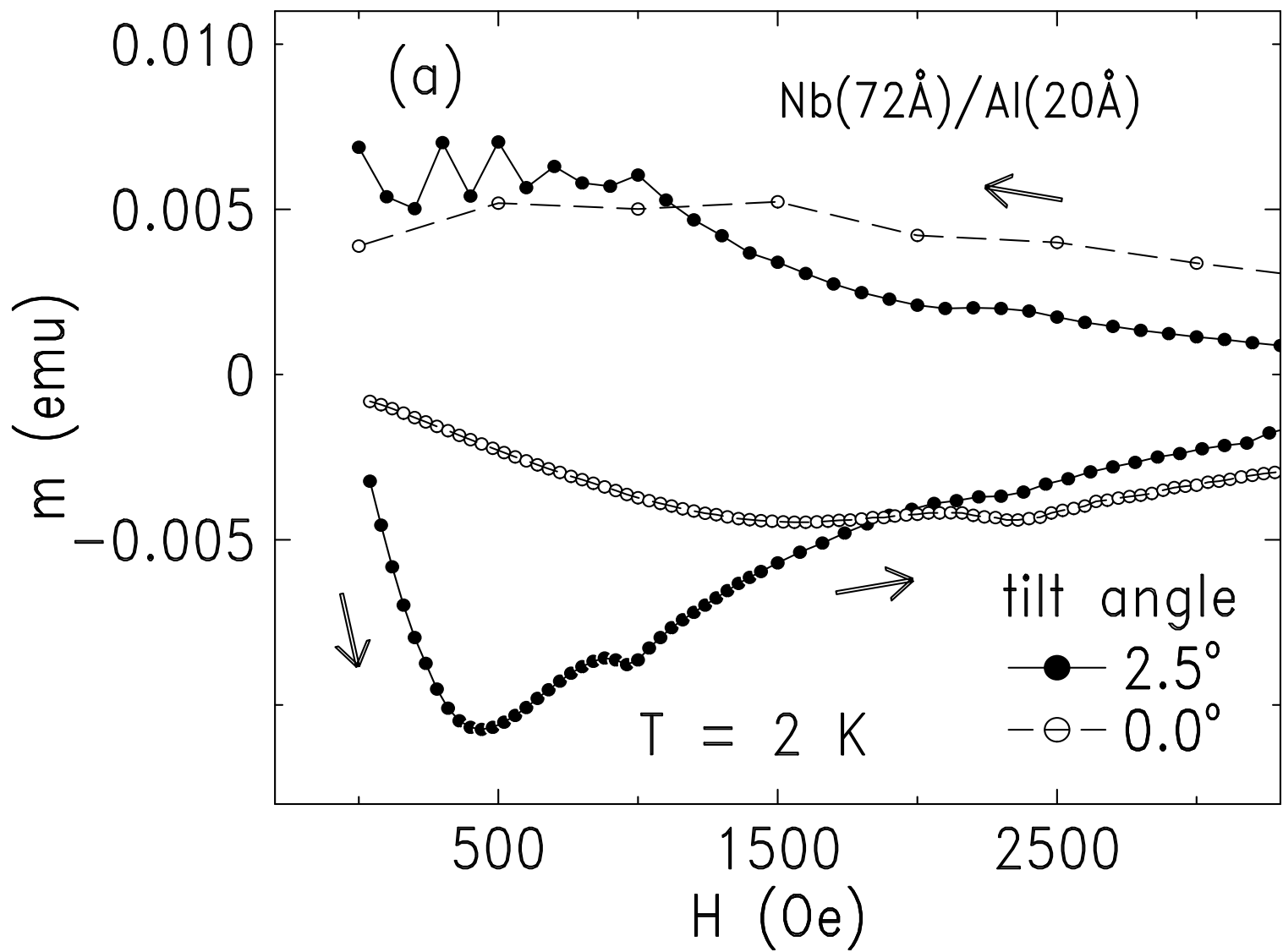
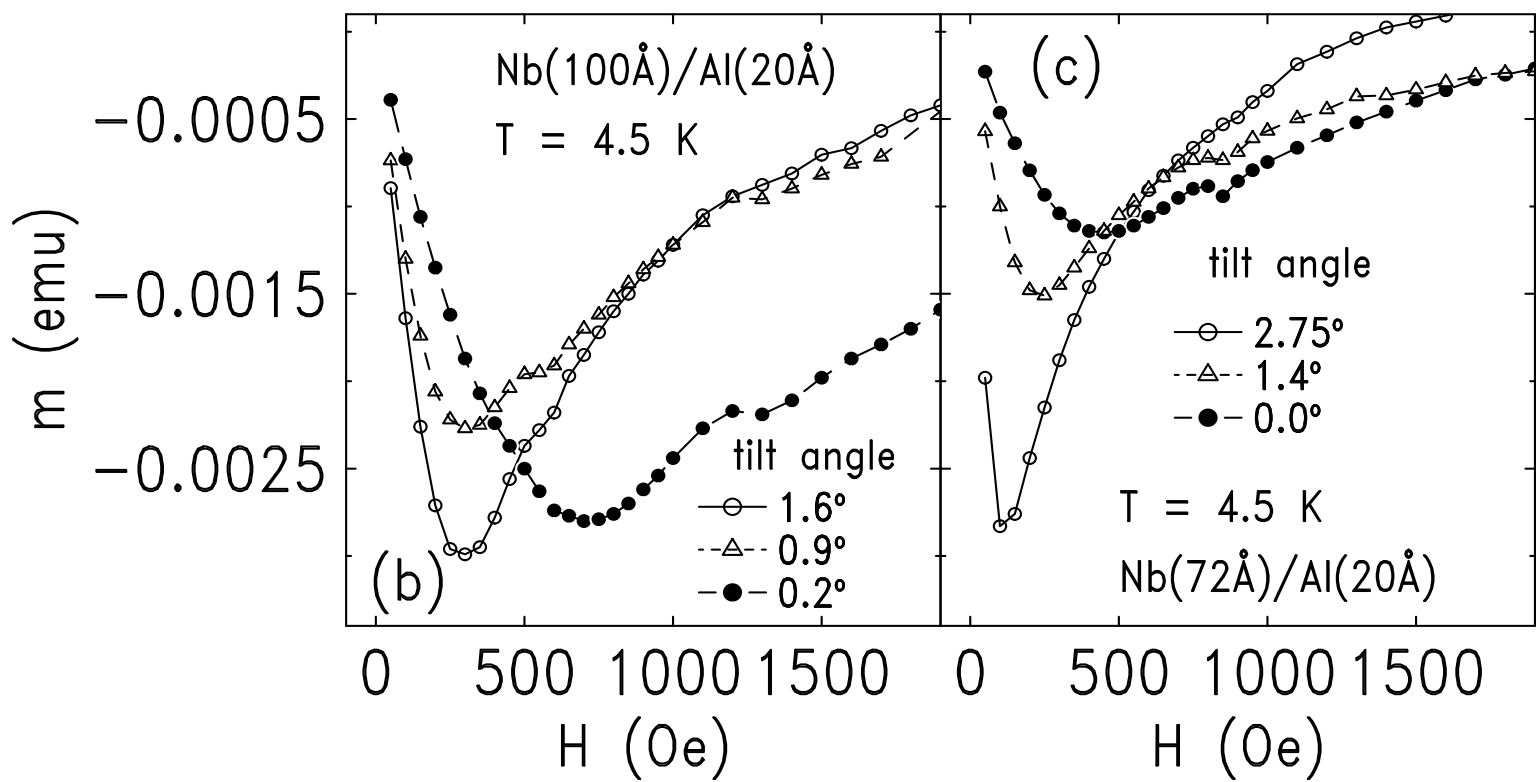
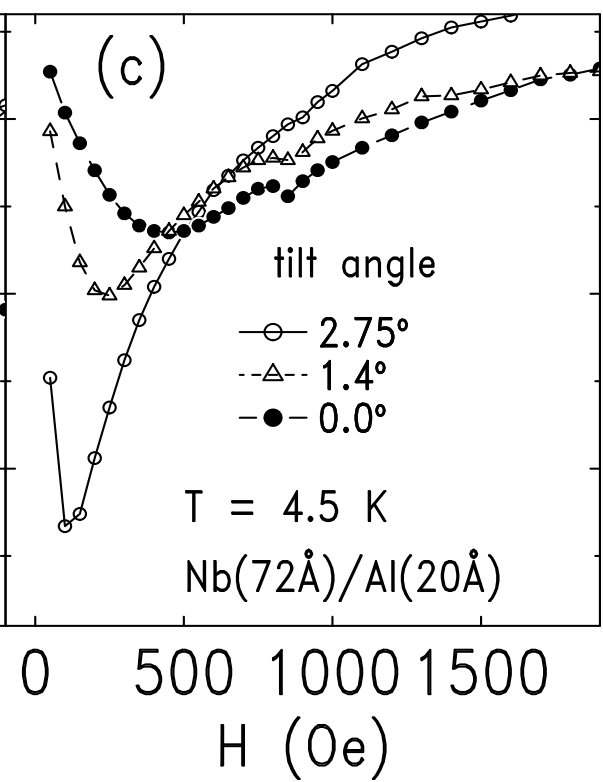
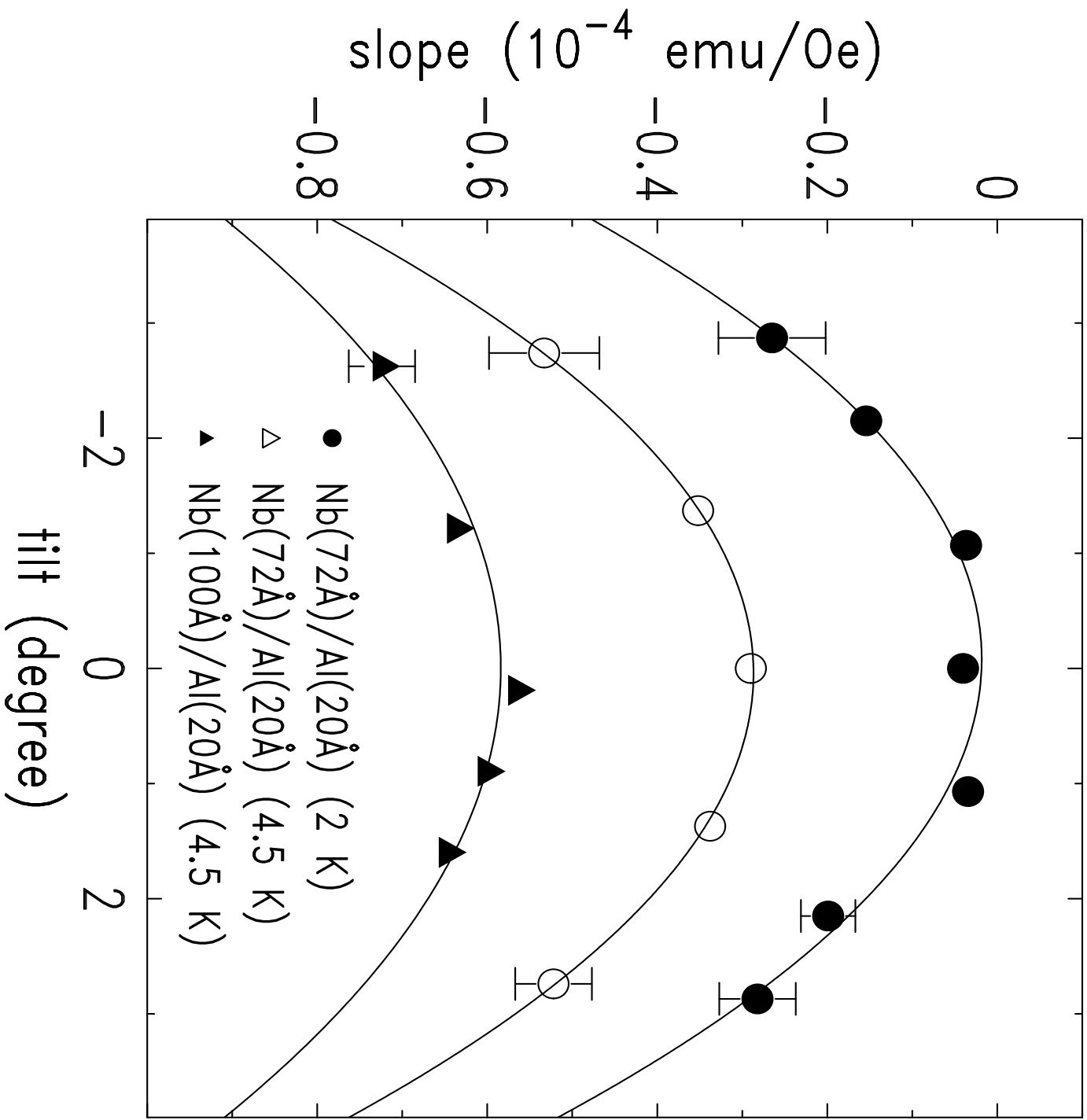
\end{document}